\documentclass[a4paper,10pt,fullpage,twocolumn]{article}
\usepackage{graphicx}
\usepackage{cite} 
\usepackage{abstract}
\newcommand{\mi}{microcalcification }
\newcommand{\mis}{microcalcifications }

\begin{document}
\twocolumn[ 
%
\title{A scalable system for microcalcification cluster automated detection in a distributed mammographic database}
\author{
Pasquale~Delogu$^{1,2}$, Maria~Evelina~Fantacci$^{1,2}$, Alessandro~Preite~Martinez$^{3}$, Alessandra~Retico$^{2}$, \\
Arnaldo~Stefanini$^{1,2}$ and Alessandro~Tata$^{2}$\\
\\
$^{1}$\small{\em Dipartimento di Fisica dell'Universit\`a di Pisa, Italy}\\
$^{2}$\small{\em Istituto Nazionale di Fisica Nucleare, Sezione di Pisa, Italy}\\
$^{3}$\small{\em Centro Studi e Ricerche Enrico Fermi, Roma, Italy}
}
\date{}
\maketitle
\begin{onecolabstract} 
A computer-aided detection (CADe) system for microcalcification cluster identification in mammograms has been developed in the framework of the EU-founded MammoGrid project. 
The CADe software is mainly based on wavelet transforms and artificial neural networks. It is able to identify microcalcifications in different datasets of mammograms (i.e. acquired with different machines and settings, digitized with different pitch and bit depth or direct digital ones).
The CADe can be remotely run from GRID-connected acquisition and annotation stations, supporting clinicians from geographically distant locations in the interpretation of mammographic data.
We report and discuss the  system performances on different datasets of mammograms and the status of the GRID-enabled CADe analysis. 
\\

\\
\end{onecolabstract}
]

\section*{Introduction}

The EU-founded MammoGrid project~\cite{MammoGrid} is currently collecting an European-distributed database of mammograms with the aim of applying the emerging GRID technologies~\cite{GRID} to support the early detection of breast cancer.
A GRID-based infrastructure would allow the resource sharing and the 
co-working between radiologists throughout the European Union. 
In this framework, epidemiological studies, tele-education of young health-care professionals, advanced image analysis and tele-diagnostic support (with and without computer-aided detection) would be enabled.

In the image processing field, we have developed and implemented in a GRID-compliant acquisition and annotation station a  computer-aided detection (CADe) system able to identify microcalcifications in different datasets of mammograms (i.e. acquired with different machines and settings, digitized with different pitch and bit depth or direct digital ones). 

This paper is structured as follows: the detection scheme is illustrated in sec.~\ref{sec:Description}, sec.~\ref{sec:MammoGridDatabase} describes the database the MammoGrid Collaboration has collected, whereas the tests carried out on different datasets of mammograms and the preliminary results obtained on a set of MammoGrid images are discussed in sec.~\ref{sec:TestRes}.

\hfill 
 
\hfill 

\section{Description of the CADe system}
\label{sec:Description}

The CADe procedure we realized is mainly based on wavelet transforms and artificial neural networks. 
Our CADe system indicates one or more suspicious areas of a mammogram where microcalcification clusters are possibly located, 
according to the following schema~\cite{Delogu}:
\begin{itemize}
\item	INPUT: digital or digitized mammogram;
\item	Pre-processing: 
	a) identification of the breast skin line and segmentation of the breast region with respect to the background; b) application of the wavelet-based filter in order to enhance the microcalcifications;
\item	Feature extraction: a) decomposition of the breast region in several $N$$\times$$N$  pixel-wide partially-overlapping sub-images to be processed each at a time;
 b) automatic extraction of the features characterizing each sub-image;
\item	Classification: assigning each processed sub-images either to the class of  microcalcification clusters or to that of normal tissue;
\item	OUTPUT: merging the contiguous or partially overlapping sub-images and visualization of the final output by drawing the contours of the suspicious areas on the original image.
\end{itemize}

\subsection{Pre-processing of the mammograms}

The  pre-processing procedure aims to enhance the signals revealing the presence of  microcalcifications, while suppressing the 
complex and noisy non-pathological breast tissue.
A mammogram is usually dominated by the low-frequency information,
whereas the \mis appear as high-frequency 
contributions.
Microcalcifications show some evident features at some
specific scales, while they are almost negligible at other scales.  
The use of the
wavelet transform~\cite{Daubechies,Meyer2,Mallat1} allows for a separation of the more important
high-resolution components of the mammogram from the less important
low-resolution ones.

Once the breast skin line is identified, the breast region is processed by the  wavelet-based filter,  according to the 
 following main steps: identification of the
family of wavelets and the level up to which the decomposition has to be performed in order to highlight the interesting details; manipulation of the wavelet coefficients (i.e. suppression of the coefficients encoding the low-frequency contributions and enhancement of those  encoding the contributions of interesting details); inverse wavelet transform. 
By properly thresholding the wavelet
coefficients at each level of the decomposition, an enhancement of the
microcalcification with respect to surrounding normal tissue can 
be achieved in the synthesized image.
In order to achieve this result, the 
wavelet basis, the level up to which
the decomposition have to be performed and the thresholding rules
to be applied to the wavelet coefficients 
have to be accurately  set. All these choices and parameters 
are application dependent. The size of the pixel pitch and 
the dynamical range of the gray level intensities
characterizing the mammograms are the most important parameters to be
taken into account.

\subsection{Feature extraction}

In order to extract from a mammogram the features to 
be submitted to the classifier,  
small regions of a mammogram are analyzed each at a time. The choice of 
fragmenting the mammogram in small sub-images is finalized  
both to reduce the amount of data 
to be analyzed at the same time and to facilitate the localization of the 
lesions possibly present on a mammogram.
The size of the sub-images  has been chosen according to  
the basic rule of considering the smallest squared area matching 
the typical size of a small  \mi cluster.  
Being the size of a single \mi rarely greater  than 1 mm, and 
 the mean  distance between two microcalcifications belonging to the same 
cluster  generally smaller than 5 mm, 
we assume a square 
with a 5 mm side to be large enough to accommodate a small cluster. 
This sub-image size is appropriate  to 
discriminate an isolated microcalcification 
(which is not considered to be a pathological sign)
from a group of \mis close together.
The length of the square side in pixel units is obviously 
determined by the pixel pitch of the digitizer or of the direct digital device.
Let us assume that our choice for the length of the square side
corresponds to $N$ pixels.
In order to avoid the accidental missing of a \mi cluster happening to be at 
the interface between two contiguous  sub-images, we use the technique of the 
partially overlapping sub-images, i.e. we let the mask for 
selecting the sub-image to be analyzed
move through the mammogram by half of the side length 
($N/2$ pixels) at each horizontal and vertical step.
In this way each region of a mammogram is analyzed more than once  
with respect to different neighboring regions.

Each  $N$$\times$$N$ pixel-wide sub-image extracted from the filtered mammogram
is processed by an
auto-associative neural network, 
used to perform an automatic
extraction of the relevant features of the sub-image.  
The implementation of an auto-associative
neural network is a neural-based method to perform an unsupervised
feature extraction~\cite{Kramer1,Kramer2,Leonard,Kuespert}. This step has been introduced in the CADe
scheme  to reduce the
dimensionality of the amount of data (the gray level intensity values of
the $N$$\times$$N$ pixels of each sub-image) to be classified by the system.
The architecture of the network we use is a bottle-neck one, consisting of three layers of
 $N^2$ input, $n$ hidden (where $n\ll N^2$)
and  $N^2$ output neurons respectively.
This neural network is trained to reproduce in output
the input values. The overall activation  of
the $n$ nodes of the bottle-neck layer summarize the relevant
features of the examined sub-image. The more the 
$N$$\times$$N$ pixel-wide sub-image  
obtained as output is close to 
the original sub-image provided as input,
the more the activation potentials of the $n$ hidden neurons 
are supposed 
to accommodate the information
contained in the original sub-image.

It is worth noticing that the implementation of an 
auto-associative neural network at this stage of the CADe scheme
allows for a strong compression of
the parameters representing each sub-image ($N^2 \to n$) 
to be passed to the following step of the
analysis.

\subsection{Classification}

We use the $n$ features extracted by the auto-associative neural
network 
to assign each sub-image   to either the class of sub-images 
containing microcalcification clusters
or the class of those consisting only of normal breast tissue.
A standard three-layer feed-forward neural network
has been chosen to perform the classification
of the $n$ features extracted from each sub-image. 
The general architecture characterizing  this net 
consists in  $n$  inputs, $h$ hidden and two output neurons, 
and the supervised training phase
is based on the back-propagation algorithm.

The performances of the training algorithm 
were evaluated according to the 5$\times$2 cross validation 
method~\cite{Dietterich}. It is the recommended test to be performed on 
algorithms that can be executed 10 times because it can provide a reliable 
estimate of the variation of the algorithm performances
due to the choice of the training set. 
This method consists in performing 5 replications of the 2-fold cross 
validation method~\cite{Stone}. 
At each replication, the available data are randomly partitioned
into 2 sets ($A_i$ and $B_i$ for $i=1,\dots5$) with an almost equal 
number of entries. 
The learning algorithm is trained on each set and
tested on the other one.
The system performances  are given 
in terms of the sensitivity and specificity values, where
the sensitivity is defined as 
the true positive fraction (fraction of malignant masses correctly classified by the system), whereas the specificity as the true negative 
fraction (fraction of benign masses correctly classified by the system). 
In order to show the trade off
between the sensitivity and the specificity, a 
Receiver Operating Characteristic (ROC) analysis has been 
performed~\cite{Metz,Hanley}. 
The ROC curve is obtained by plotting the  
true positive fraction
versus the false positive fraction 
of the cases (1 - specificity), computed  
while the decision threshold of the classifier is varied.
Each decision threshold results in a corresponding
operating point on the curve.

\section{The MammoGrid distributed database}
\label{sec:MammoGridDatabase}

One of the main goals of the EU-founded MammoGrid project is the realization of a GRID-enabled European database of mammogram, with the aim of supporting the collaboration among clinicians from different locations in the analysis of mammographic data.
Mammograms  in the DICOM~\cite{DICOM} format are collected
through the MammoGrid acquisition and annotation workstations installed in the participating hospitals. Standardized images are stored into the GRID-connected  database. The image standardization is realized by the Standard-Mammogram-Form (SMF) algorithm~\cite{SMF} developed by the Mirada Solutions Company$^{\rm TM}$, a partner of the MammoGrid project.
The SMF provides a normalized representation of the mammogram, i.e. independent of the data source and of the acquisition technical  parameters (e.g. mAs, kVp and breast thickness).

The dataset of fully-annotated mammogram containing microcalcification clusters available at present to CADe developers is constituted  by 
 123 mammograms belonging to 57 patients: 46 of them have been collected and digitized at the University Hospital of Udine (IT), whereas the remaining 11 were acquired by the full-field digital mammography system GE Senographe 2000D at the Torino Hospital (IT); all have been stored in the MammoGrid database by means of the MammoGrid workstation prototype installed in Udine.

\section{Tests and results}
\label{sec:TestRes}

As the amount of mammograms  collected at present in the MammoGrid database is too small for properly training the neural networks implemented in the characterization and classification procedures of our CADe, we used a larger dataset of mammograms for developing the system. Once the CADe has been trained and tested, we adapted it to the MammoGrid images and we evaluated its performances on the MammoGrid database. 
The dataset used for training and testing the CADe was extracted from the fully-annotated MAGIC-5 database~\cite{magic5}. We used 375 mammograms containing microcalcification clusters  and 610 normal mammograms
digitized with a pixel pitch of 85 $\mu$m and an effective dynamical range of 12 bit per pixel.

\subsection{Training and testing the CADe on the MAGIC-5 database}

\begin{figure}
\centering
\includegraphics[width=3.5in]{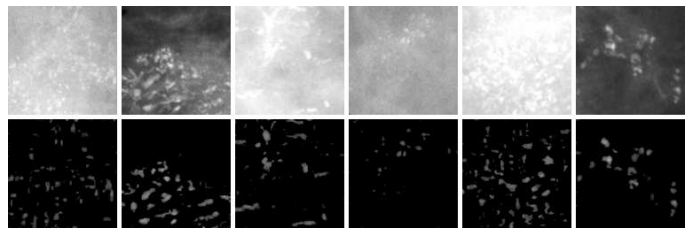}
\caption{Examples of the wavelet-based filter  performances on tissues with different densities (top/bottom: original/filtered sub-images containing microcalcification clusters).}
\label{fig:filtered-im}
\end{figure}

To perform the  multi-resolution analysis we considered the 
Daubechies family of wavelet~\cite{Daubechies}, in particular the
db$5$ mother wavelet.
The decomposition is performed up to the
 forth level. 
We found out that the
resolution level 1 mainly shows  the high-frequency noise included in
the mammogram, whereas the levels 2, 3 and 4 contain  the high-frequency
components related to the presence of microcalcifications.
Levels greater than 4 exhibit a strong correlation with larger structures 
possibly present in the normal breast tissue.
In order to enhance microcalcifications, 
the approximation coefficients at level 4
and the detail coefficients at the first level 
were neglected. By contrast, the statistical analysis of the distributions of 
the remaining detail coefficients lead us to keep into account 
for the synthesis procedure
only those coefficients whose values exceed $2 \sigma$, where $\sigma$
is the standard deviation of the coefficient distribution at that level.
Some examples of the performance of the filter on mammographic images
containing
\mi clusters embedded in tissues with different densities are shown in fig.~\ref{fig:filtered-im}. 

\begin{figure}
\centering
\includegraphics[width=3.5in]{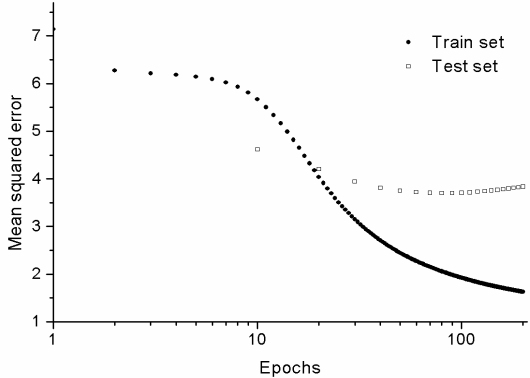}
\caption{Mean squared errors on the train and test sets in the learning phase of the 
auto-associative neural network: the minimum error on the test set is reached 
between 80 and 90 epochs.}
\label{fig:earlystop}
\end{figure}

The training and testing of the auto-associative neural network has been performed on a dataset of 149 mammograms containing microcalcification clusters  and 299 normal mammograms. 
The size $N$ of the sub-images to be analyzed by this  neural network has been chosen as  
 $N=60$, thus corresponding to a physical region of 5.1$\times$5.1 mm$^2$.
The number $n$  of units in the hidden layer has been fixed according to the 
requirement of having the minimum number of neurons
allowing for a good generalization capability of the system.
Assigning too much neurons to the hidden layer would 
facilitate the convergence of the learning phase, 
but  it could reduce the  generalization  capability
of the network. Moreover, a too populated hidden layer could 
set too stringent limits on the minimum number of patterns needed for  
training the neural classifier implemented 
in the following step of the analysis.
By contrast, a too small hidden layer would lead to the 
 saturation of some of the hidden units and thus negatively affect
the overall performance of the system.
A good compromise between these two opposite trends has been reached by 
assigning 80 units to the hidden layer. 
The network architecture is thus fixed to be: 
3600 input, 80 hidden and 3600 output
neurons.  
The algorithm used in the training procedure is the standard
back-propagation with momentum and the activation function is a
sigmoid.  We used a learning rate of 0.4 and a momentum of 0.2.
The behavior of the mean squared error computed during the learning 
procedure at each epoch on the train set and every ten epochs on the test
set is shown in fig.~\ref{fig:earlystop}.
The training phase has been stopped once 
the error on the test
set has reached the minimum value (early stop). As shown in  fig.~\ref{fig:earlystop} it 
happens between epochs 80 and 90. The training phase was thus forced to 
finish in 85 epochs.

The dataset used for the supervised training of the feed-forward
neural classifier is constituted by 156 mammograms with 
\mi clusters  and 
241 normal mammograms. 
The standard back-propagation
algorithm was implemented and the best performance 
 were achieved with 10 neurons in the hidden layer.
The performances our learning algorithm achieved according to the 5$\times$2 cross-validation method are reported 
in tab.~\ref{tab:5X2}
in terms of the sensitivity and specificity values.
\begin{table}[t]
\caption{Evaluation of the performances of the standard back-propagation learning algorithm for the neural classifier according to the 5$\times$2 cross validation method.}
\label{tab:5X2} 
\centering
 \begin{tabular}{cccc}  \hline 
Train Set & Test Set & Sensitivity (\%) & Specificity (\%)\\
\hline 
$A_1$ & $B_1$ &	94.4	& 91.8 \\
$B_1$ & $A_1$ &	92.8	& 91.1\\
$A_2$ & $B_2$ &	92.3	& 90.9\\
$B_2$ & $A_2$ &	93.4	& 92.0\\
$A_3$ & $B_3$ &	92.0	& 90.5\\
$B_3$ & $A_3$ &	94.5	& 91.6\\
$A_4$ & $B_4$ &	92.9	& 93.9\\
$B_4$ & $A_4$ & 94.2	& 93.0\\
$A_5$ & $B_5$ &	94.6	& 91.5\\
$B_5$ & $A_5$ &	93.0	& 91.7\\
\hline 
\end{tabular}
\end{table}
As can be noticed, the performances the neural classifier achieves are robust, 
i.e. almost 
independent of the 
partitioning of the available data into the train and test sets. 
The average performances achieved in the testing phase 
are  93.4\% for the sensitivity and 91.8\% for the specificity.

\begin{figure}
\centering
\includegraphics[width=3.5in]{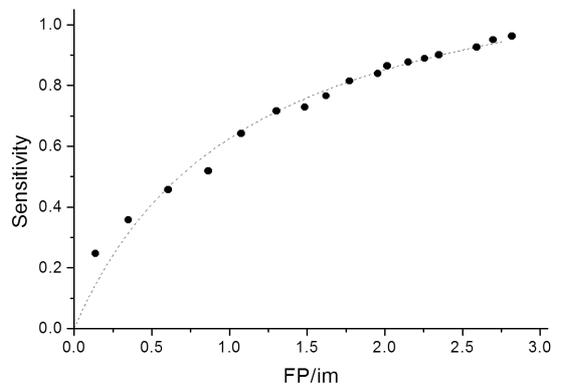}
\caption{FROC curve obtained on a test set of 140 mammograms (70 containing 89 \mis clusters and 70 
normal views) extracted from the MAGIC-5 database.}
\label{fig:FROC}
\end{figure}

Once each sub-image of 
a mammogram has been assigned a degree of suspiciousness,  
the contiguous or partially-overlapping suspicious sub-images 
have to be merged in order to evaluate the system performances 
on the entire mammographic images.
A cluster detection criterion has to be {\it a priori} defined.
The effect the choice of the detection criteria in addition to the
size of the annotated region has on the CAD performance evaluation have been  
systematically examined in the literature~\cite{Brake, Kallergi}.
As there is no universal scoring method currently in use for evaluating 
the performances of a CAD system for \mi cluster detection,  we briefly describe the  detection
criteria we adopted:
\begin{itemize}
\item a true cluster is considered  detected if 
the region indicated by the system includes 
two or more \mis located within the associated truth  circle;
\item all findings outside the truth
circle are considered as false positive (FP) detections. 
\end{itemize}
The  CADe performances were globally evaluated on a test set of 140 images of the MAGIC-5 database
(70 with microcalcification clusters and 70 normal images) in terms of the free-response operating 
characteristic  (FROC) analysis~\cite{Chakraborty} (see  fig.~\ref{fig:FROC}). 
The FROC curve is obtained by plotting the sensitivity of the system versus 
the number of  FP detection per image (FP/im), 
while the decision threshold of the classifier is varied.
In particular, as shown in the figure, a sensitivity value of 88\% 
is obtained at a rate of 2.15 FP/im.

\subsection{Testing the CADe on the MammoGrid database}

The CADe system we developed and tested on the MAGIC-5 database has been adapted to the MammoGrid SMF images by using the following procedure:
\begin{itemize}
\item the wavelet-based filter has been tuned on the SMF mammograms;
\item the remaining steps of the analysis, i.e. the neural-based characterization and classification of the sub-images  have been directly imported from the MAGIC-5 CADe software. 
\end{itemize}
According to the MammoGrid project work-flow~\cite{MammoGrid},  the CADe algorithm has to run on mammograms previously processed 
by the SMF software~\cite{SMF}.
The SMF mammograms  are characterized by a different pixel pitch (100 $\mu$m instead of 85 $\mu$m) and a different effective dynamical range (16 bit per pixel instead of 12) with respect to the MAGIC-5 mammograms.  
A \mi digitized with a $85 \mu$m pixel pitch scanner appears bigger 
(in pixel units) with respect to the same object digitized with a 
$100 \mu$m pixel pitch. Therefore,  the filter to be applied to the  MammoGrid
mammograms is required to be sensitive to smaller object. 
A different choice in the range of scales to be considered in the
analysis has proved to be comfortable
for accommodating the difference in the pixel pitch.  
Once the matching of the effective dynamical ranges of the two databases has been performed, 
the wavelet decomposition
is performed up to level 3 instead of 4, being the
details at level 4 too big to be correlated to microcalcifications.
Only the details at levels 2 and 3 (exceeding $2 \sigma$ of the
experimental distribution) are kept into account for the synthesis.
A test of this scaling procedure has been performed on the mammograms of 15 patients acquired  both by the 
MAGIC-5 and by the MammoGrid acquisition workstations.
\begin{figure}
\centering
\includegraphics[width=3.5in]{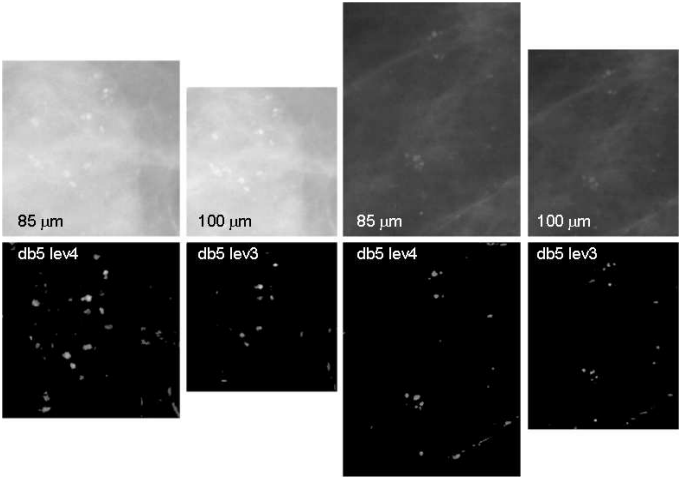}
\caption{Examples of the performances of the scaling procedure for the CADe filter.}
\label{fig:scaling}
\end{figure}
As shown in fig.~\ref{fig:scaling} 
the matching of the  dynamical ranges and the scaling of the wavelet-analysis parameters allows the CADe filter to generate very similar processed images.

The performances the rescaled CADe achieves on the images of the MammoGrid database are the following:
a sensitivity of 82.2\% is obtained at a rate of 4.15 FP/im.  
If the analysis is performed independently on the digitized and on the direct digital images, the results are:
a 82.1\% sensitivity at a rate of 4.8 FP/im in the first case, whereas a 
 83.3\% sensitivity at a rate of 1.6 FP/im in the second case.
As can be noticed, the number of FP detection per image in the case of digitized images is appreciably higher with respect to the corresponding  rate for the directly digital images. 
Despite the SMF algorithm performs a sort of normalization of images acquired in different conditions, the digitized images are intrinsically noisier.     
A comparison with the FROC obtained on the MAGIC-5 database reported in fig.~\ref{fig:FROC} points out that 
the overall CADe system performances in the case of the  MammoGrid database are not as good as those obtained 
on the  MAGIC-5 dataset.
One possible explanation for this decrease in sensitivity, is that the MammoGrid database contains already a large number of non-easily detectable cases. In this case an improvement of the CADe performances would be achieved once the database is enlarged. 

\section{Conclusion}
We developed a CADe system for microcalcification cluster identification suitable for different sets of data (digitized or direct digital, acquired with different acquisition parameters, etc.).
This CADe system has been developed and tested on the MAGIC-5 database and then adapted to the MammoGrid database of SMF mammograms by re-scaling some of the wavelet-filter parameters. 
This choice is  motivated by two main reasons:  the amount of fully-annotated SMF images containing microcalcification clusters available at present to the MammoGrid CADe developers is not large enough to perform a new training of the neural networks implemented  in the characterization and classification procedures;
moreover, the visual aspect of the filtered sub-images in the case both of  MAGIC-5 images and SMF images is actually very similar. This makes us confident that the generalization capability of the  neural networks would account for  the difference in resolution of the two original datasets.
The scaling procedure we developed  has two main advantages:
the wavelet filter is the only part of the analysis one has to 
tune on the characteristics of a new dataset, whereas 
the neural-based characterization and classification procedures do no need to be modified; 
this scalable system  can be tested even on very small databases not allowing for the learning procedure of the neural networks to be properly carried out.  

The preliminary results obtained on MammoGrid database are encouraging. Once the planned increase in the population of the database is realized, a complete and more robust test of the CADe performance on the pan-European MammoGrid database would be carried out. 

The CADe software is currently available on the  GRID-connected acquisition and annotation workstation prototypes
installed  in the Hospitals of the MammoGrid Consortium. 
The CADe can be remotely executed  on the distributed database and the clinical evaluation
of the CADe as second reader of screening mammograms has already started.

\section*{Acknowledgments}
 This work has been partially supported by the EU-founded MammoGrid project and by the Istituto Nazionale di Fisica Nucleare, 
Sezione di Pisa, Italy.

\end{document}